\DocumentMetadata{testphase=new-or-1}

\documentclass[sn-nature,Numbered,iicol]{sn-jnl}

\usepackage{graphicx}%
\usepackage{multirow}%
\usepackage{amsmath,amssymb,amsfonts}%
\usepackage{amsthm}%
\usepackage{mathrsfs}%
\usepackage[title]{appendix}%
\usepackage{xcolor}%
\usepackage{textcomp}%
\usepackage{manyfoot}%
\usepackage{booktabs}%
\usepackage{algorithm}%
\usepackage{algorithmicx}%
\usepackage{algpseudocode}%
\usepackage{listings}%
\usepackage{multirow}

\usepackage[utf8]{inputenc} 
\usepackage[T1]{fontenc}    
\usepackage{hyperref}       
\usepackage{url}            
\usepackage{nicefrac}       
\usepackage{microtype}      
\usepackage{lipsum}		
\usepackage{doi}
\usepackage{color,soul}
\usepackage{caption}
\usepackage{booktabs}
\usepackage{placeins}
\usepackage{tikz}
\usepackage{subcaption}
\usepackage{comment}
\usepackage[printonlyused, nohyperlinks]{acronym}
\usepackage{mleftright}
\usepackage{tabularx}

\newcommand{\tikzcircle}[2][red,fill=red]{\tikz[baseline=-0.5ex]\draw[#1,radius=#2] (0,0) circle ;}%

\newcommand\blfootnote[1]{%
  \begin{NoHyper}
  \renewcommand\thefootnote{}\footnote{#1}%
  \addtocounter{footnote}{-1}%
  \end{NoHyper}
}

\begin{document}

\raggedbottom

\title{Virchow: A Million-Slide Digital Pathology Foundation Model}

\author[1]{\fnm{Eugene} \sur{Vorontsov}\textsuperscript{\textdagger}}
\author[1]{\fnm{Alican} \sur{Bozkurt}\textsuperscript{\textdagger}}
\author[1]{\fnm{Adam} \sur{Casson}\textsuperscript{\textdagger}}
\author[1]{\fnm{George} \sur{Shaikovski}\textsuperscript{\textdagger}}
\author[1]{\fnm{Michal} \sur{Zelechowski}\textsuperscript{\textdagger}}
\author[1]{\fnm{Siqi} \sur{Liu}\textsuperscript{\textdagger\textdaggerdbl}}
\author[2]{\fnm{Kristen} \sur{Severson}}
\author[2]{\fnm{Eric} \sur{Zimmermann}}
\author[2]{\fnm{James} \sur{Hall}}
\author[2]{\fnm{Neil} \sur{Tenenholtz}}
\author[2]{\fnm{Nicolo} \sur{Fusi}}
\author[1]{\fnm{Philippe} \sur{Mathieu}}
\author[1]{\fnm{Alexander} \sur{van Eck}}
\author[1]{\fnm{Donghun} \sur{Lee}}
\author[1]{\fnm{Julian} \sur{Viret}}
\author[1]{\fnm{Eric} \sur{Robert}}
\author[1]{\fnm{Yi Kan} \sur{Wang}}
\author[1]{\fnm{Jeremy D.} \sur{Kunz}}
\author[1]{\fnm{Matthew C. H.} \sur{Lee}}
\author[1]{\fnm{Jan H.} \sur{Bernhard}}
\author[1]{\fnm{Ran A.} \sur{Godrich}}
\author[1]{\fnm{Gerard} \sur{Oakley}}
\author[3]{\fnm{Ewan} \sur{Millar}}
\author[1,4]{\fnm{Matthew} \sur{Hanna}}
\author[1]{\fnm{Juan} \sur{Retamero}}
\author[1]{\fnm{William A.} \sur{Moye}}
\author[1]{\fnm{Razik} \sur{Yousfi}}
\author[1,5]{\fnm{Christopher} \sur{Kanan}}
\author[1,4]{\fnm{David} \sur{Klimstra}}
\author[1]{\fnm{Brandon} \sur{Rothrock}}
\author[1]{\fnm{Thomas J.} \sur{Fuchs}}
\affil[1]{\orgname{Paige}, \city{NYC}, \state{NY} \country{United States}}
\affil[2]{\orgname{Microsoft Research}, \city{Cambridge}, \state{MA} \country{United States}}
\affil[3]{\orgdiv{NSW Health Pathology}, \orgname{St George Hospital}, \city{Sydney} \country{Australia}}
\affil[4]{\orgname{Memorial Sloan Kettering Cancer Center}, \city{NYC}, \state{NY} \country{United States}}
\affil[5]{\orgname{University of Rochester}, \city{Rochester}, \state{NY} \country{United States}}

\abstract{
The use of artificial intelligence to enable precision medicine and decision support systems through the analysis of pathology images has the potential to revolutionize the diagnosis and treatment of cancer. Such applications will depend on models' abilities to capture the diverse patterns observed in pathology images. To address this challenge, we present Virchow, a foundation model for computational pathology. Using self-supervised learning empowered by the DINOv2 algorithm, Virchow is a vision transformer model with 632 million parameters trained on 1.5 million hematoxylin and eosin stained whole slide images from diverse tissue and specimen types, which is orders of magnitude more data than previous works. The Virchow model enables the development of a pan-cancer detection system with 0.949 overall specimen-level \ac{AUROC} across 17 different cancer types, while also achieving 0.937 \ac{AUROC} on 7 rare cancer types. The Virchow model sets the state-of-the-art on the internal and external image tile level benchmarks and slide level biomarker prediction tasks. The gains in performance highlight the importance of training on massive pathology image datasets, suggesting scaling up the data and network architecture can improve the accuracy for many high-impact computational pathology applications where limited amounts of training data are available.
}

\maketitle
\def\thefootnote{\textdaggerdbl}\footnotetext{ Corresponding author. siqi.liu AT paige DOT ai\\}\def\thefootnote{\arabic{footnote}}

\def\thefootnote{\textdagger}\footnotetext{These authors contributed equally to this work.\\}\def\thefootnote{\arabic{footnote}}

\blfootnote{This is a live paper that will be updated with results from ongoing work.}

\keywords{Foundation model \and Self-supervised \and Pathology \and Whole slide image \and Representation learning}

\section{Main}
\label{sec:introduction}

\begin{figure*}
    \centering
    \includegraphics[width=\linewidth]{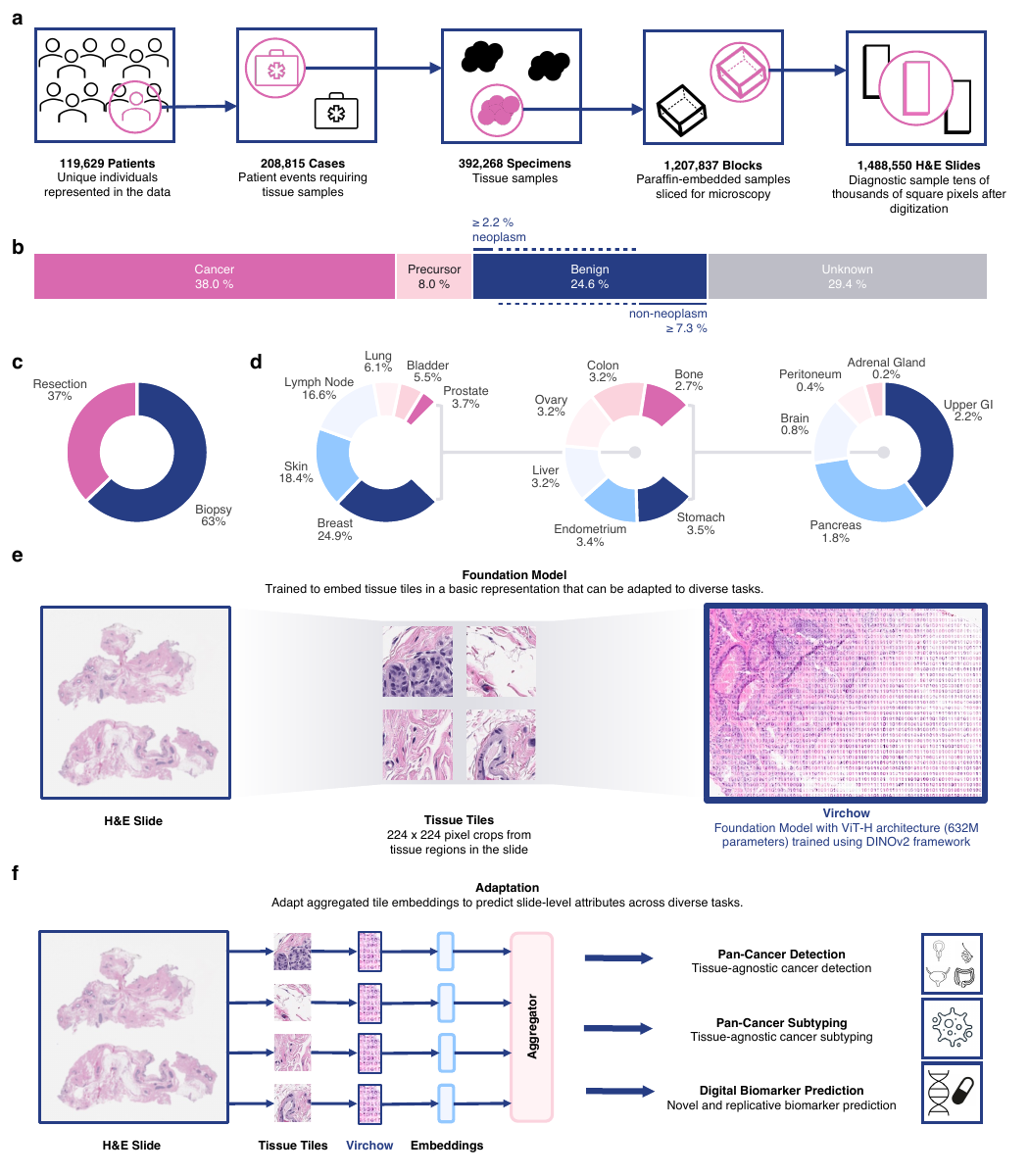}
    \caption{Overview of the training dataset (\textbf{a}-\textbf{d}), training algorithm (\textbf{e}), and application (\textbf{f}) of Virchow, a foundation model for digital pathology. \textbf{a.} The training data can be described in terms of patients, cases, specimens, blocks or slides as shown. (\textbf{b}-\textbf{d}) The slide distribution as a function of cancer status (\textbf{b}), surgery (\textbf{c}), and tissue type (\textbf{d}). \textbf{e.} The dataflow during training which requires processing the slide into tiles, which are then cropped into global and local views. \textbf{f.} Schematic of applications of the foundation model using an aggregator model to predict attributes at the slide level.}
    \label{fig:teaser}
\end{figure*}

Pathology is essential for the diagnosis and treatment of cancer. As pathology data is not natively digital, for many decades, the field has remained relatively unchanged. With the rise of digitization of \ac{HE} stained microscopy slides, also known as \acp{WSI}, a new field of computational pathology~\cite{deng2020deep, srinidhi2021deep, cooper2023machine, song2023artificial} is emerging. Computational pathology applies \ac{AI} to digitized \acp{WSI} to support the diagnosis, characterization, and understanding of disease~\cite{fuchs2011computational, abels2019computational}. Initial work has focused on clinical decision support tools to enhance current workflows~\cite{bejnordi2017diagnostic, raciti2022clinical, da2021independent, perincheri2021independent, raciti2020novel, campanella2019clinical}. However given the incredible gains in performance of computer vision, a sub-field of artificial intelligence focused on images, more recent studies~\citep{cdh1_Reis-Filho_Pareja, wagner2023fully, coudray2018classification, kather2019deep, bilal2021development} attempt to unlock new insights from routine \acp{WSI} and reveal undiscovered outcomes such as therapeutic response~\citep{xie2022computational_response}. If successful, such efforts would enhance the utility of \ac{HE}-stained \acp{WSI} and reduce reliance on  specialized and often expensive \ac{IHC} or genomic testing~\citep{kacew}.

A major factor in the performance gains of computer vision models has been the creation of very large deep neural networks, termed \textit{foundation models}. Foundation models are trained on enormous datasets using a family of algorithms, referred to as self-supervised learning (e.g.~\cite{chen2020simple, zhou2021ibot, caron2020unsupervised, caron2021emerging, he2022masked}), which do not require task-specific, curated labels. Foundation models generate data representations, known as embeddings, that can generalize well to a variety of downstream tasks~\citep{bommasani2021opportunities}. These properties make foundation models well-suited to the pathology domain given the increasing volume of unlabeled data and diverse \ac{AI} applications, including cancer detection, cancer subtyping, biomarker quantification, mitotic event counting, and survival prediction. A successful pathology foundation model would capture a broad spectrum of patterns, including cellular morphology, tissue architecture, staining characteristics, nuclear atypia, mitotic figures, necrosis, inflammatory response, neovascularization, texture features, and biomarker expression and therefore would be well-suited to predicting a wide-variety of \ac{WSI} characteristics.

Foundation model performance crucially depends on dataset and model size as demonstrated by scaling law results~\cite{kaplan2020scaling, zhai2022scaling, openai_gpt-4_2023}. Modern foundation models in the natural image domain use millions of images (e.g. ImageNet~\cite{deng2009imagenet}, JFT-300M~\cite{sun2017revisiting} and LVD-142M~\cite{oquab2023dinov2}) to train models with hundreds of millions to billions of parameters~\cite{dosovitskiy2020image}. Datasets of this scale are challenging to collect in the medical domain due to the frequency of image acquisition and challenges in sharing data between institutions. Most of the proposed foundation models in computational pathology~\cite{wang2022transformer, ciga2022self, filiot2023scaling, azizi2023robust, kang2023benchmarking} primarily leverage \ac{TCGA}~\cite{weinstein2013cancer}, an open-access repository of approximately 29k \ac{HE}-stained whole slide images from 32 cancer types, and employ architectures with fewer than 100M parameters (see Sec.~\ref{app:hist_FM} for detailed discussion of models). Three recent works leverage larger, proprietary datasets: (1) 400k \acp{WSI} corresponding to 77k patients from Mount Sinai Health System~\cite{campanella2023computational}, (2) 100k \acp{WSI} from Massachusetts General Hospital, Brigham \& Women's Hospital and the Genotype-Tissue Expression consortium~\cite{chen_general-purpose_2023} and (3) 100k \acp{WSI} combining proprietary data and \ac{TCGA}~\cite{dippel2024rudolfv}. These works scale the model size to 22M~\cite{campanella2023computational} and 307M~\citep{chen_general-purpose_2023, dippel2024rudolfv} parameters. From these recent works, it is evident that pathology image features produced with self-supervised learning by early-stage foundation models outperform image features trained on natural images, and that this performance improves with dataset and model scale.

We present the first million-scale pathology foundation model, \textit{Virchow}, named in honor of Rudolf Virchow\footnote{Rudolf Virchow (pronounced vir-kov) is the father of modern pathology~\citep{schultz2008rudolf,reese1998fundamentals} and proposed the first theory of cellular pathology~\citep{virchow1860cellular}}. Virchow is trained on data from approximately 100 thousand patients corresponding to approximately 1.5 million \ac{HE} stained \acp{WSI} acquired from \ac{MSKCC} which is at least an order of magnitude larger than prior training datasets in pathology (detailed in Fig.~\ref{fig:teaser}a and Sec.~\ref{sec:training-dataset}). The training data is composed of cancerous and benign tissue, collected via biopsy (63\%) and resection (37\%), from 17 high-level tissue types (Fig.~\ref{fig:teaser}b,c,d). Virchow, a 632M parameter vision transformer model, is trained using the DINOv2 algorithm~\cite{oquab2023dinov2}, a multi-view student-teacher self-supervised algorithm (Figure~\ref{fig:teaser}e, see Sec.~\ref{sec:dino_description} for training details). DINOv2 leverages global and local regions of tissue tiles to learn to produce embeddings of \ac{WSI} tiles (Fig.~\ref{fig:teaser}e), which can be aggregated across slides and used to train a variety of downstream predictive tasks (Fig.~\ref{fig:teaser}f).

We implemented a wide array of benchmarks to evaluate the performance of Virchow embeddings on the downstream computational pathology tasks. Virchow consistently outperforms competing models on all benchmarks. Motivated by highlighting the potential clinical impact of the foundation model, we assess the performance of models trained using the Virchow embeddings to predict specimen-level cancer across different organs. Virchow embeddings are shown to outperform or match all the baseline models on all tested cancer types, notably including rare cancers and \ac{OOD} data. Similarly, Virchow embeddings yield state-of-the-art biomarker prediction performance. Our results demonstrate that large-scale foundation models can be the basis for robust results in a new frontier of computational pathology.

\section{Results}
\label{sec:experiments}

We evaluated the Virchow model embeddings on two categories of slide-level computational pathology applications: pan-cancer detection (Sec.~\ref{sec:pancancer_results}) and biomarker prediction (Sec.~\ref{sec:biomarker}). These benchmarks require training a weakly supervised aggregator model to transfer tile embeddings to slide-level predictions. We also performed a series of tile-level linear probing benchmarks to directly assess the embeddings on individual tissue tiles (Sec.~\ref{sec:results_tile}).

\subsection{Pan-cancer detection}
\label{sec:pancancer_results}

\begin{figure*}
    \centering
    \includegraphics[width=\linewidth]{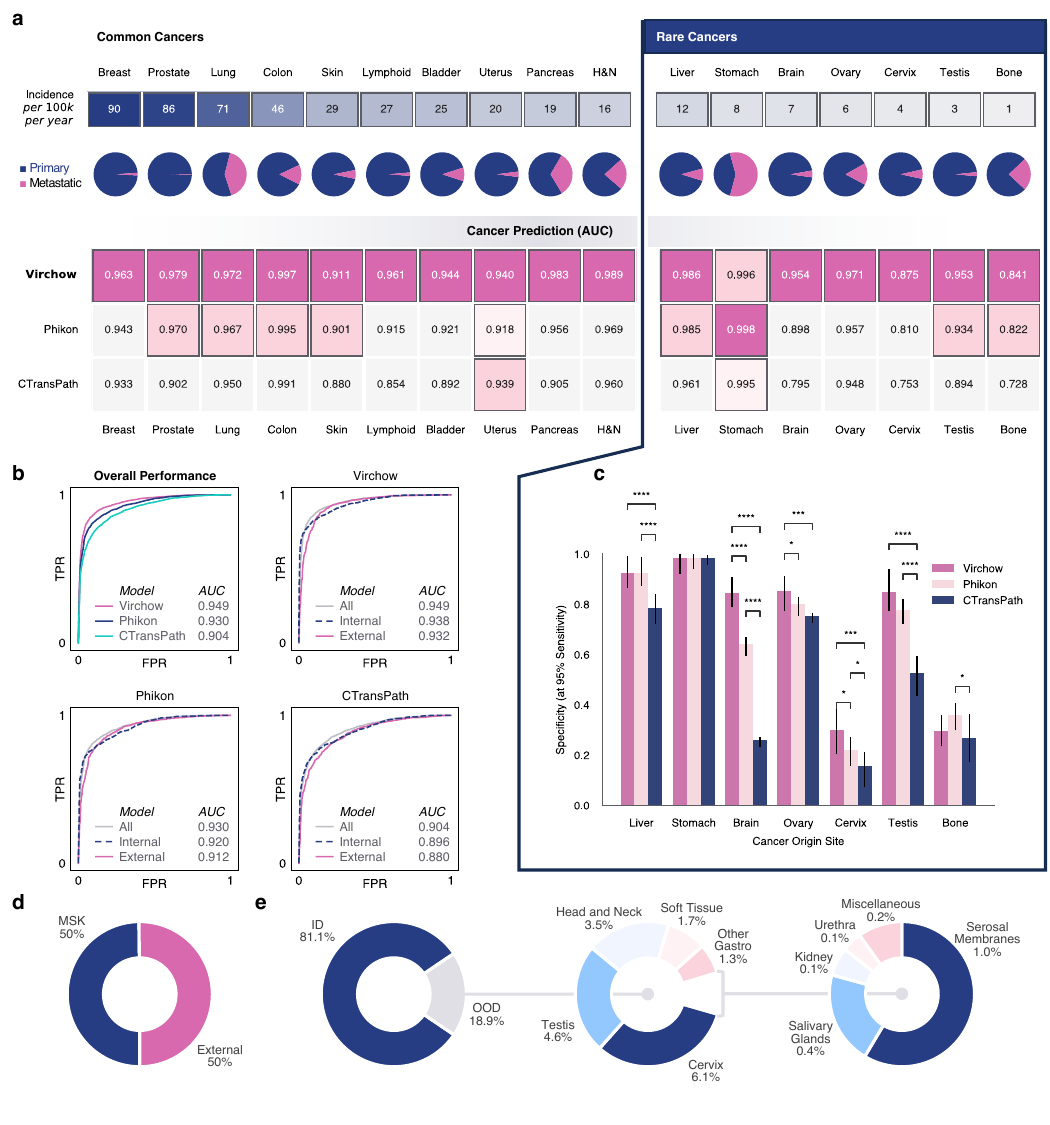}
    \caption{Pan-cancer detection results (\textbf{a}-\textbf{c}). Detection is specimen-level, produced with an aggregator network trained on Virchow, Phikon, or CTransPath tile embeddings. \textbf{a.} Cancer prediction performance (\ac{AUROC}) stratified by cancer type as determined by origin tissue (``H\&N'' is head and neck). The incidence rate of each cancer is shown. Virchow embeddings enable the best cancer detection performance across all cancer types and performance remains robust on rare cancers. For each cancer type, the \ac{AUROC} corresponding to the statistically significantly (p < 0.05) top performing embeddings is highlighted in magenta. When more than one \ac{AUROC} is not gray, performance is ``tied” (no statistically significant difference) \textbf{b.} \Ac{ROC} curves showing the overall pan-cancer detection performance, as well as performance stratified across internal \ac{MSKCC} data vs. data coming from diverse external institutions. All evaluation data is withheld from training. \textbf{c.} Sensitivity at 95\% specificity for rare cancer detection (* p < 0.05, ** p < 0.01, *** p < 0.001, **** p < 0.0001). \textbf{d.} Half of the specimens come from diverse external institutions (OOD data). \textbf{e.} ID vs. OOD tissues in the evaluation dataset. Some of the OOD tissues arise from cancer metastases.
}
    \label{fig:pancancer}
\end{figure*}

A key aim of our work was to develop a single model to detect cancer, especially rare cancer, across various tissues. The proposed pan-cancer detection system predicts the presence of cancer using foundation model embeddings as input (see Sec.~\ref{sec:methods_pancancer} for architecture and training algorithm details). The pan-cancer detection model is evaluated on slides from \ac{MSKCC} as well as slides submitted for consultation to \ac{MSKCC} from numerous external sites globally. Model performance was stratified across 10 common and 7 rare cancer types (see Sec.~\ref{sec:methods_pancancer} for a detailed breakdown of the evaluation dataset).

Embeddings generated by the proposed Virchow model, Phikon~\citep{filiot2023scaling}, and CTransPath~\citep{wang2022transformer} are evaluated (see Sec.~\ref{sec:embeddings} for further detail on embeddings). Pan-cancer aggregators are trained using specimen-level labels, maintaining the same training protocol for all embeddings (see Sec.~\ref{sec:methods_pancancer} for details).

Virchow embeddings yielded the best cancer detection performance on all cancer types (Fig.~\ref{fig:pancancer}a). Pancancer detection using Phikon embeddings achieved statistically similar performance (p < 0.05) for 5 of the 10 common cancer types and 4 of the 7 rare cancer types; nevertheless, in all but one case, the specific \ac{AUROC} score was lower. Overall the pan-cancer model achieved an \ac{AUROC} of 0.949 with Virchow embeddings, 0.930 with Phikon embeddings, and 0.904 with CTransPath embeddings (Fig.~\ref{fig:pancancer}b; all significantly different with p << 0.001).

Rare cancer detection performance is particularly noteworthy. Compared to the aforementioned \ac{AUROC} of 0.949 overall, Virchow embeddings yielded an \ac{AUROC} of 0.937 on rare cancers, demonstrating robust generalization to rare data. Performance across the individual rare cancers was however non-uniform with detection of cervical and bone cancers proving more challending (\ac{AUROC} < 0.9), irrespective of the embeddings used (Fig.~\ref{fig:pancancer}b,c). Virchow embeddings improved cervix detection to 0.875 \ac{AUROC} compared with 0.810 when using Phikon embeddings or 0.753 when using CTransPath embeddings. Similarly, Virchow embeddings yielded  0.841 \ac{AUROC} for bone cancer detection, compared to 0.822 with Phikon and 0.728 with CTransPath. Finally, using Virchow embeddings yielded the greatest improvement for detecting brain cancer, producing an \ac{AUROC} of 0.954, compared to 0.898 or 0.795 with Phikon and CTransPath, respectively. 

The pan-cancer models demonstrated robustness to out-of-distribution data when tested on data external to \ac{MSKCC}, regardless of the choice of embeddings. Detection performance dropped from 0.938, 0.920, and 0.896 \ac{AUROC} on internal \ac{MSKCC} data by 0.006, 0.008, and 0.016, respectively for Virchow, Phikon, and CTransPath embeddings (Fig.~\ref{fig:pancancer}b). The performance drop is minor and expected as both the Virchow foundation model training set and the pan-cancer model training set contained only data internal to \ac{MSKCC}. On the other hand, half of the specimens in the evaluation set are sourced externally from \ac{MSKCC} (Fig.~\ref{fig:pancancer}d).

In addition to external sites, the evaluation dataset contains \ac{OOD} tissues that were not seen during model training. These comprise 18.9\% of the specimens in the dataset, as shown in Fig.~\ref{fig:pancancer}e. Overall, pancancer detection generalizes across cancer types, including rare cancers, as well as on \ac{OOD} data when using foundation model embeddings.

\subsection{Biomarker detection}
\label{sec:biomarker}

\begin{table*}[!t]
\centering
\begin{tabular}{@{}rccc@{}}
\toprule
   & \multicolumn{3}{c}{ \acs{AUROC} (2.5\% CI, 97.5\% CI)}  \\ \cmidrule{2-4}
         Backbone  & ColonMSI            & BladderFGFR         & LungEGFR            
                                          \\ \midrule
CTransPath \cite{wang2022transformer} & 0.970 (0.946, 0.989) & 0.882 (0.824, 0.930) & 0.807 (0.758, 0.852) \\
Phikon \cite{filiot2023scaling}     & 0.957 (0.905, 0.992) & 0.886 (0.838, 0.930) & 0.821 (0.771, 0.864) \\
\textbf{Virchow} & \textbf{0.972} (0.950, 0.989) & \textbf{0.902} (0.862, 0.941) & \textbf{0.853} (0.804, 0.891) \\ \bottomrule
\end{tabular}
\caption{Case-level \ac{AUROC} scores on the testing sets for different biomarker targets using the
aggregator network trained on tile-level embeddings from the baseline backbones and
Virchow.}
\label{tab:biomarker_results}
\end{table*}

Biomarker prediction using standard \ac{HE} stained images represents another significant use of computational pathology. We train one aggregator network for each biomarker using the foundation model embeddings to predict the presence of the biomarker. Specifically, models are trained to predict colon \ac{MSI}, 
bladder \ac{FGFR}, and lung \ac{EGFR}. These biomarkers play a crucial role in the diagnosis and treatment of various cancers and each is described in further detail in Sec.~\ref{sec:methods_biomarker}.
Samples included in the biomarker detection datasets had been previously subjected
to targeted sequencing using the FDA-authorized \ac{MSK-IMPACT} assay. \Acp{WSI} from the histological sections matching the respective blocks
utilized for DNA extraction and \ac{MSK-IMPACT} sequencing \citep{zehir_mutational_2017} were utilized. \ac{MSK-IMPACT} targeted sequencing data was analyzed to determine the status of genetic alterations and establish a binary label indicating the presence or absence of the variants, i.e. the biomarker. Similarly to the pan-cancer evaluation, the publicly available Phikon model~\citep{filiot2023scaling} and CTransPath model~\citep{wang2022transformer} were used as baseline models for comparisons. 

The overall \ac{AUROC} scores are shown in Tab.~\ref{tab:biomarker_results}.
Additionally, to provide a more comprehensive statistical analysis, we have included the 2.5 percentile and 97.5 percentile confidence intervals, which were obtained using 1000 bootstrapping iterations. 
The Virchow model demonstrates a consistently high performance across all three biomarkers, with \ac{AUROC} scores of 0.972 (95\% CI: 0.950, 0.989) for ColonMSI, 0.902 (95\% CI: 0.862, 0.941) for BladderFGFR, and 0.853 (95\% CI: 0.804, 0.891) for LungEGFR.  The table underscores the superior performance of the Virchow model in the context of this digital biomarker prediction across different tissues.

\subsection{Tile-level benchmarks}
\label{sec:results_tile}

\begin{figure*}
    \centering
    \includegraphics[width=\linewidth]{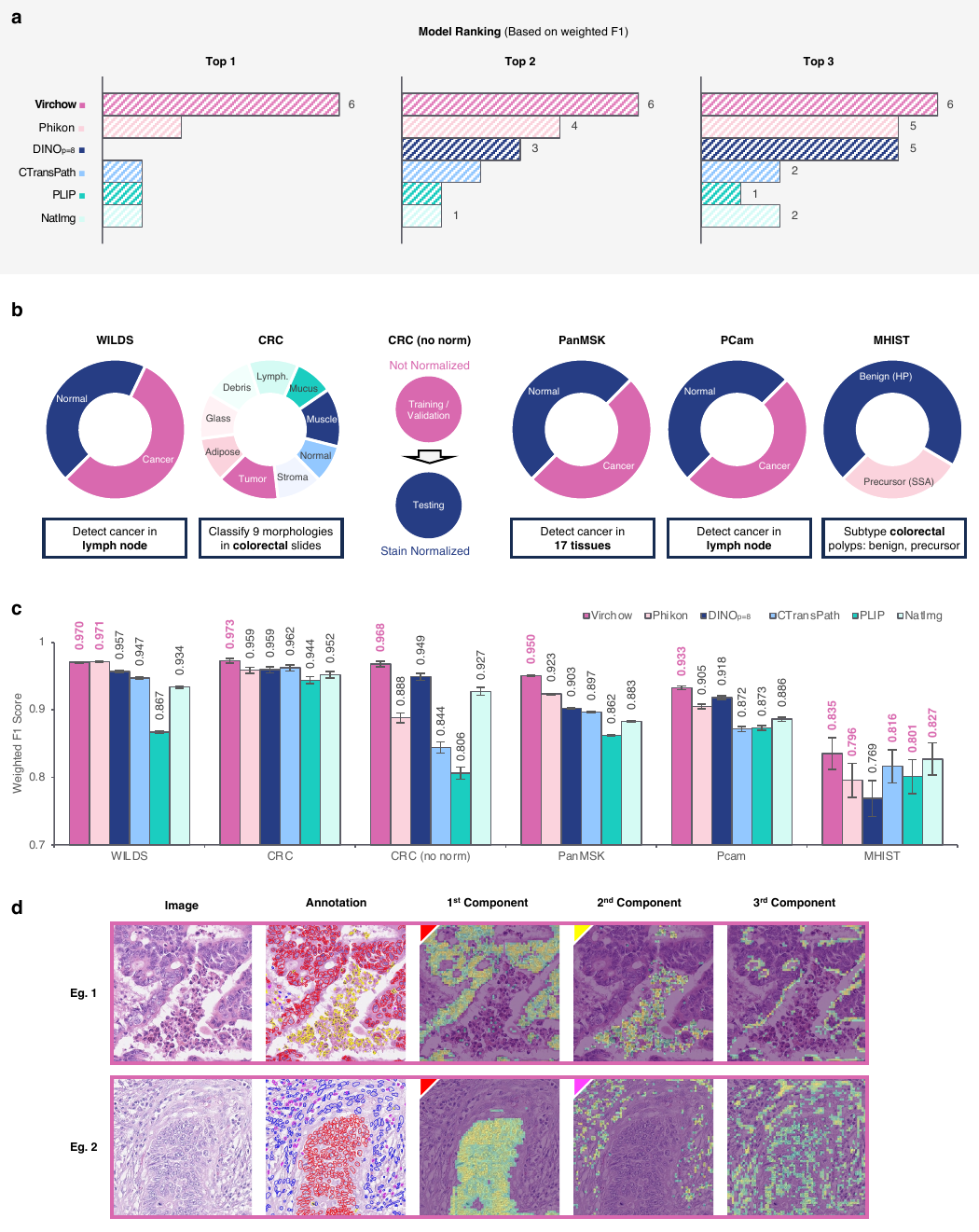}
    \caption{A summary of tile-level linear probing. \textbf{a.} The number of tasks in which each model scored in the top-x. \textbf{b.} A description of each task. \textbf{c}. The weighted F1 score for each of the six models and six tasks. \textbf{d}.
    Virchow discovers cells in the \acs{consep} dataset: malignant epithelium (\tikzcircle[red, fill=red]{3pt} red), miscellaneous (\tikzcircle[yellow, fill=yellow]{3pt} yellow), and inflammatory (\tikzcircle[magenta, fill=magenta]{3pt} magenta) cells.
    }
    \label{fig:linear_probing}
\end{figure*}

We use the tile-level benchmarks to assess the robustness and generalizability of the foundation model embeddings on WSI tissue tiles directly. 
These tasks are evaluated using linear probing of the tile embeddings. We therefore compare Virchow embeddings to baseline model embeddings by applying the same linear probing protocol for each model, using the same training, validation, and testing data splits (see Sec.~\ref{sec:methods_tile} for further details). Analysis is performed both on public datasets and on the internal \ac{MSKCC} pan-cancer dataset.

The internal multi-tissue dataset for pan-cancer detection at the tile level (referred to as PanMSK) is an \ac{ID} benchmark, as it is composed from annotations on a held-out set of patients across the entire diverse set of tissue groups selected for training (Fig.~\ref{fig:teaser}d). The public datasets are \acf{OOD} benchmarks. These include multi-class colorectal cancer classification (NCT-CRC-HE-100K and NCT-CRC-HE-100K-NONORM,~\citep{kather_jakob_nikolas_2018_1214456,kather2019predicting}), colorectal polyp classification (MHIST,~\citep{wei2021petri}), and breast lymph node cancer classification (\ac{PCam}, \citep{veeling2018rotation,bejnordi2017diagnostic}).

In addition to  Phikon~\citep{filiot2023scaling} and CTransPath~\citep{wang2022transformer}, $\text{DINO}_{p=8}$~\cite{kang2023benchmarking} (49M parameter model trained using \ac{TCGA} and an internal dataset), PLIP~\citep{huang2023visual} (87M parameter model trained using pathology image-text pairs), and NatImg~\citep{oquab2023dinov2} (1.1B parameter model trained on 142 million natural images) are evaluated.

As shown in Fig.~\ref{fig:linear_probing}a,c,  Virchow matches or surpasses baseline performance across all datasets containing different tissue types and cancer subtypes (Fig.~\ref{fig:linear_probing}a-c; see Tab.~\ref{tab:tile_bench} for additional metrics). As shown in Fig.~\ref{fig:linear_probing}a, Virchow consistently places in the top-1 by weighted F1 score across all six tasks, demonstrating robust performance on diverse tasks. The closest competing models are Phikon and $\text{DINO}_{p=8}$, with Phikon tying in top-1 twice and scoring in the top-2 results four times, and $\text{DINO}_{p=8}$ scoring in among the top-2 three times. Virchow demonstrates strong \ac{OOD} performance on the WILDS and ``CRC (no norm)'' tasks. The WILDS test data is sourced from a hospital that is not encountered in the training set. The ``CRC (no norm)'' task introduces a distribution shift from the stain-normalized training set by avoiding stain normalization on the testing set. Without normalization, Virchow's performance declines by only $-0.005$ in weighted F1 score.  This indicates that Virchow is robust to variations in data preprocessing.

To qualitatively evaluate whether the embeddings learnt by Virchow tend to separate the image into semantically meaningful clusters of features, we performed an unsupervised feature analysis similar to the procedure in \cite{oquab2023dinov2}. We visualized the embedded feature separation on the \ac{consep} dataset~\citep{graham2019hover} of \ac{HE} stained slides with colorectal adenocarcinoma (detailed in Sec.~\ref{sec:methods_tile_pca}).

We observe approximate semantic segmentation of the cell types in the \ac{consep} images (Fig.~\ref{fig:linear_probing}d). In both examples, the first principal component highlighted malignant epithelium (\tikzcircle[red, fill=red]{3pt} red) cells. The second principal component respectively highlighted miscellaneous cells (\tikzcircle[yellow, fill=yellow]{3pt} yellow) and inflammatory (\tikzcircle[magenta, fill=magenta]{3pt} magenta) cells. DINOv2 was shown to learn a similar semantic feature separation on natural images, allowing foreground/background separation (e.g. discriminating a bus or a bird from the background) as well as part annotation (e.g. wheels vs. windows in a bus)~\citep{oquab2023dinov2}. Here, we show that this emerging property of the model carries over to the pathology domain. This encouraging result supports our expectation that the unsupervised features learnt by Virchow are meaningful and interpretable for a wide range of downstream tasks.

\section{Discussion}
\label{sec:discussion}

The field of computational pathology achieved a major milestone following the successful application of \ac{MIL}~\cite{campanella2019clinical, ilse2018attention}. Using \ac{MIL} with labels at the level of groups of slides has enabled clinically relevant diagnostics by scaling to training datasets on the order of ten thousand \acp{WSI}~\cite{raciti2022clinical, da2021independent, perincheri2021independent, raciti2020novel, campanella2019clinical}. These early works typically initialized the model's embedding parameters using pre-trained model weights, often those trained on ImageNet in a supervised setting. This process, called transfer learning, was motivated by the observation that model performance critically depends on the model's ability to capture image features. In-domain transfer learning was not possible given the limited availability of labeled datasets.
 Now self-supervised learning is enabling in-domain transfer by removing the label requirement, driving a second wave of scaling to tens of thousands of \acp{WSI} to inform image representation~\cite{wang2022transformer, ciga2022self, chen2022scaling, filiot2023scaling, azizi2023robust,kang2023benchmarking}.
Virchow marks a major increase in training data scale to 1.5 million \acp{WSI} -- a volume of data that is over three thousand times the size of ImageNet~\cite{deng2009imagenet} as measured by the total number of pixels. This large scale of data in turn motivates large models which can capture the diversity of image features in \acp{WSI}. In this work we have demonstrated the value of this approach.

There are many design choices made during foundation model training that merit further discussion. The impact of the particular self-supervised learning algorithm on model performance remains an open question. A central characteristic of histopathology data is the long tailed distribution of interesting features, both in terms of pathologies and basic tissue features. This can be perceived as a ``class'' imbalance between benign and pathological features and across cancer types. In the natural image domain, class imbalance has been shown to produce poor representations with contrastive self-supervised learning~\citep{assran2022hidden}. While DINOv2 can be viewed as primarily a mean teacher based method, it also includes a contrastive regularizer. Nevertheless, many recent works, in addition to ours, choose this method~\cite{chen_general-purpose_2023, dippel2024rudolfv, filiot2023scaling}. Indeed, prior work found there was no clear best performing approach when training four different self-supervised algorithms on 37 thousand \acp{WSI} \citep{kang2023benchmarking}, although the DINO~\cite{caron2021emerging} approach most often had the best performance. While both mean teacher based methods like DINOv2 and reconstruction based methods such as the \ac{MAE}~\citep{he2022masked} perform well on class-imbalanced data, the embeddings produced by the latter yield worse linear probing performance and require an additional finetuning step~\citep{he2022masked,shekhar2023objectives}. 

Another set of open questions pertain to the importance of making domain-specific design and training decisions as opposed to leveraging techniques demonstrated in the natural image domain. Overall, results do suggest there is value in training on histopathology data as opposed to foundation models trained using natural images as was demonstrated in the tile-level results (note that due to the model size, the natural image model, NatImg, is a particularly strong non-pathology natural image baseline). However beyond the training data, there are more subtle ways in which one might hope to tailor methods to histopathology. For instance, self-supervised learning techniques augment the brightness, contrast and color of images. Several works~\cite{tellez2019quantifying, ciga2022self, gullapally2023synthetic} have investigated the impact of color augmentation motivated by stain variation, the effect of differences in staining protocol and scanner type which do not reflect underlying differences in pathology. All of the aforementioned studies have demonstrated improvements in performance using color augmentation and several models have employed domain-orientated approaches~\cite{wang2022transformer, dippel2024rudolfv}. Digitized \acp{WSI} are typically stored at fixed magnifications, e.g. 5$\times$, 10$\times$, 20$\times$, and have significantly less variability in object scale than natural images. This feature also draws into question augmentation protocols which typically crop and resize images. In addition to color variation, Ciga et al.~\cite{ciga2022self} investigated the impact of random cropping on model performance and found less random cropping generally improved performance although the largest observed delta for any setting was 5\%. The impact of using cropping parameters from the natural image literature as in this work and Filiot et al.~\cite{filiot2023scaling} as opposed to less resizing as in Chen et al.~\cite{chen_general-purpose_2023} and Ciga et al.~\cite{ciga2022self} is unclear.

Our work has several limitations. The training dataset is acquired  from one center with limited scanner types. As with most histopathology foundation models, embeddings are generated at the tile level as opposed to the slide level and therefore require training an aggregation model. A deep investigation of aggregator architectures and training procedures is beyond the scope of this work. As is the case for all models aiming for clinical application, thorough stratified performance validation is required. 

Although we trained both Virchow and the pan-cancer detection aggregator on only data internal to \ac{MSKCC}, we demonstrate that pan-cancer prediction remains robust on data from external sites, on \ac{OOD} tissue types, and on rare cancer types. Rare cancers are important and make up about 25\% of the data. Virchow also outperformed other embeddings for biomarker prediction, a type of task with limited data that benefits from the expressiveness of a large foundation model. Similarly, our technical tile-level benchmarks perform well on all \ac{OOD} data. Virchow embeddings outperform those of smaller-scale foundation models on all tasks, demonstrating the performance and robustness that can be gained from learning a rich representation of the diversity of pathology images, at scale.

\section{Methods}
\label{sec:methods}

\subsection{Million-scale training dataset}
\label{sec:training-dataset}
The dataset used for self-supervised training was sourced from \ac{MSKCC}. It is comprised of 1,488,550 \acp{WSI} derived from 119,629 patients. These \acp{WSI} are all stained with \ac{HE}, a routine stain that stains the nuclei blue and the extracellular matrix and cytoplasm pink. The \acp{WSI} are scanned at 20$\times$ resolution, or 0.5 \ac{mpp}, using Leica scanners. 17 high-level tissue groups are included, as illustrated in Fig.~\ref{fig:teaser}c. 

 \acp{WSI} are gigapixel in size and are cumbersome to use directly during training. Instead Virchow was trained on tissue tiles that were sampled from foreground tissue in each \ac{WSI}. To detect foreground, each \ac{WSI} was downsampled 16$\times$ with bilinear interpolation and every pixel of the downsampled image was considered as tissue if its hue, saturation, and value were within [90, 180], [8, 255], and [103, 255], respectively. All non-overlapping 224 $\times$ 244 tiles containing at least 25\% tissue by area were collected.

\subsection{Virchow architecture and training}
\label{sec:dino_description}
Virchow employs the \ac{ViT} ``huge'' architecture (\ac{ViT}-H/14), a vision transformer~\citep{dosovitskiy2020image} with 632 million parameters, and was trained using the DINOv2~\citep{oquab2023dinov2} self-supervised learning algorithm as illustrated in Fig.~\ref{fig:dinov2}. 
The Vision Transformer (ViT) is an adaptation of the transformer model for image analysis, treating an image as a sequence of patches. These patches are embedded and processed through a transformer encoder that uses self-attention mechanisms. This approach allows ViT to capture complex spatial relationships across the image.
DINOv2 is based on a student-teacher paradigm: given a student network and a teacher network, each using the same architecture, the student is trained to match the representation of the teacher. The student network is information-limited, as it is trained using noisy variations of input tiles. The teacher network is a slowly updated \ac{EMA} of past student networks; matching the teacher achieves an effect similar to ensembling over prior student predictions~\citep{tarvainen2017mean}. The student learns a global representation of an image by matching the teacher's class token, as well as local representations by matching the teacher's patch tokens. Patch tokens are only matched for a select subset of tokens that are randomly masked out of an input image (for the student), as done in masked image modeling~\citep{xie2022simmim}. Additional regularization helps DINOv2 trained models outperform the earlier DINO variant~\cite{caron2021emerging}.

The default hyperparameters for training the DINOv2 model were used for Virchow, as detailed in~\cite{oquab2023dinov2} with the following changes: a learning rate warmup of 495,000 iterations (instead of 100,000) and a teacher temperature schedule of 0.04 to 0.07 in 186,000 iterations. Virchow was trained using AdamW ($\beta_1=0.9$, $\beta_2=0.999$) with float16 precision. Note that with \ac{ViT}-H, we used 131,072 prototypes (and thus 131,072-dimensional projection heads).

During distributed training, each minibatch was sampled by randomly selecting one \ac{WSI} per GPU and 256 foreground tiles per \ac{WSI}.

\subsection{Embeddings}
\label{sec:embeddings}

For a 224$\times$224 input tile image, we define a Virchow embedding as the concatenation of the class token and the mean across all 256 of the other predicted tokens. This produces an embedding size of 2,560 (1,280 $\times$ 2). For Phikon, we use only the class token, as recommended by~\cite{filiot2023scaling}. For CTransPath, we use the mean of all tokens as there is no class token.

\subsection{Pan-cancer detection}
\label{sec:methods_pancancer}
Specimen-level pan-cancer detection requires a model which aggregates foundation model embeddings from all foreground tiles of all \acp{WSI} in a specimen to detect the presence of cancer. All pan-cancer detection models trained in this work use an Agata~\citep{raciti2022clinical} aggregator model, weakly supervised with multiple instance learning (see Sec.~\ref{sec:agata-details} for architecture details).

\textbf{Training data} To train the aggregator model, we prepared a subset of the training dataset used for training Virchow (Sec.~\ref{sec:training-dataset}), combined with specimen-level labels (block-level for prostate tissue) indicating the presence or absence of cancer extracted from synoptic and diagnostic reports. The training and validation datasets combined consist of 177,742 slides across 47,839 specimens. 

\textbf{Aggregator training} We trained the Agata aggregator, as specified in Sec.~\ref{sec:agata-details}. Since the label is at the level of the specimen, all tiles belong to the same specimen need to be aggregated during training. Training using embeddings for all tiles of a specimen is prohibitively memory-intensive. We thus select the slide with the highest predicted cancer probability per specimen and backpropagate the gradients only for that slide.

As baselines, we also trained aggregators using Phikon and CTransPath embeddings. All aggregators were trained for 25 epochs using the cross-entropy loss and the AdamW optimizer with a base learning rate of 0.0003. During each training run, the checkpoint with the highest validation \ac{AUROC} was selected for evaluation. 

\textbf{Testing dataset} The pan-cancer detection models are evaluated on a combination of data sourced from \ac{MSKCC} and external institutions. None of the patients in the evaluation set were seen during training. The dataset contains 23,408 slides from 6372 specimens across 17 high-level tissue types. We hypothesise that the more data the foundation model is trained on, the better the downstream task performance, especially on data-constrained tasks. In order to test this hypothesis, we categorize cancer types into common or rare cancer groups. According to the National Cancer Institute, rare cancers are defined as those occurring in fewer than 15 people out of 100 thousand each year in the United States~\citep{rare_cancers_def}. Based on this definition, common cancer comprises 14,610 slides  from 3770 specimens originating in breast, prostate, lung, colon, skin, lymph nodes, bladder, uterus, pancreas, and \ac{HN}; whereas rare cancer comprises 8798 slides from 2602 specimens originating in liver, stomach, brain, ovary, cervix, and testis. Note that each cancer type is determined by its tissue of origin and thus may appear as primary cancer in the same tissue or as metastatic cancer in any other tissue. On the other hand, benign specimens for each cancer type were sampled only from the tissue of origin. Figure~\ref{fig:pancancer}a shows the distribution between primary and metastatic for each cancer type.

The testing dataset includes 15,941 slides from 3175 specimens collected at \ac{MSKCC} (denoted as ``Internal'' in Fig.~\ref{fig:pancancer}b, in addition to 7467 slides (3197 specimens) sent to \ac{MSKCC} from institutions around the world (``External'' in Fig.~\ref{fig:pancancer}b). 

\textbf{Label extraction} To establish the clinical cancer diagnosis at the specimen level, a rule-based natural language processing system was employed. This system decomposes case-level reports to the specimen level and analyzes the associated clinical reports with each specimen, thereby providing a comprehensive understanding of each case.

\textbf{Metrics} The performance of the three models is compared using 2 metrics: \ac{AUROC} and specificity at $95\%$ sensitivity. For \ac{AUROC}, the pair-wise DeLong's test~\citep{delong1988comparing} with Holm's method~\citep{holm_bonferroni} for correction is applied to check for statistical significance. For specificity, first Cochran's Q test~\citep{cochran_q} is applied, and then McNemar's test~\citep{mcnemar1947note} is applied post-hoc for all pairs with Holm's method for correction. The confidence interval in Fig.~\ref{fig:pancancer}c are calculated using Wilson's method~\cite{wilson1927probable}. In addition to overall analysis, stratified analysis is also conducted for each cancer type.

\subsection{Biomarker detection}
\label{sec:methods_biomarker}
We formulated each biomarker prediction task as a binary pathology case classification problem, where a positive label indicates the presence of the biomarker. Each case consists of one or more \ac{HE} slides that share the same binary label. We randomly split each dataset into training, validation, and testing subsets, ensuring no patient overlap, as shown in Tab.~\ref{tab:biomarker-data-stats}. The clinical importance of each biomarker is described below.

\begin{table*}
\centering
\begin{tabular}{@{}rrrrrrrrrr@{}}
\toprule
           & \multicolumn{3}{c}{ColonMSI} & \multicolumn{3}{c}{BladderFGFR} & \multicolumn{3}{c}{LungEGFR} \\
        \cmidrule(r{2pt}){2-4} \cmidrule(l{2pt}){5-7} \cmidrule(l{2pt}){8-10}
Split      & Cases  & Slides & Positive ratio & Cases & Slides & Positive ratio &  Cases  & Slides & Positive ratio \\
 \midrule
Training      & 2,029   & 2,291   & 0.10          & 520    & 542    & 0.24          & 2,186  & 2,858   & 0.28          \\
Validation & 334    & 384    & 0.12          & 259    & 275    & 0.29          & 356   & 457    & 0.29          \\
Testing       & 335    & 373    & 0.13          & 259    & 270    & 0.25          & 358   & 457    & 0.28          \\
\bottomrule
\end{tabular}
\caption{Statistics of the case-level biomarker target datasets, including the number of cases, the number of slides, and the proportion of positive labels.}
\label{tab:biomarker-data-stats}
\end{table*}

\textbf{Colon-MSI} \Acf{MSI} occurs when DNA regions with short, repeated sequences (microsatellites) are disrupted by single nucleotide mutations, leading to variation in these sequences across cells. Normally, \ac{MMR} genes (MSH1, MSH2, MSH6, PMS2) correct these mutations, maintaining consistency in microsatellites. However, inactivation of any \ac{MMR} gene (through germline mutation, somatic mutation, or epigenetic silencing) results in an increased rate of uncorrected mutations across the genome. \ac{MSI} is detected using \ac{PCR} or next-generation sequencing, which identifies a high number of unrepaired mutations in microsatellites, indicative of \ac{dMMR}. \Ac{MSI-H} suggests dMMR in cells, identifiable via \ac{IHC}, which shows absent staining for \ac{MMR} proteins. \Ac{MSI-H} is present in approximately 15\% of \acp{CRC}, often linked to germline mutations that elevate hereditary cancer risk. Consequently, routine \ac{MSI} or \ac{IHC}-based \ac{dMMR} screening is recommended for all primary colorectal carcinoma samples. The ColonMSI dataset, comprising 2,698 CRC samples with 288 \ac{MSI-H}/\ac{dMMR} positive cases, uses both \ac{IHC} and \ac{MSK-IMPACT} sequencing for \ac{dMMR} and \ac{MSI-H} detection, prioritizing \ac{IHC} results when both test outcomes are available.

\textbf{Bladder-FGFR} alterations screening in bladder carcinoma allows the identification of patients targetable by \ac{FGFR} inhibitors. Anecdotal experience from pathologists suggested there may be a morphologic signal for \ac{FGFR} alterations~\citep{al-ahmadie_somatic_2011}.
The FGFR3 binary label focuses on FGFR3 p.S249C, p.R248C, p.Y373C mutations and FGFR3-TACC3 fusions based on data from the \ac{MSK-IMPACT} cohort. From the total of 1051 \acp{WSI}, 25.8\% have FGFR3 alterations.

\textbf{Lung-EGFR}  oncogenic mutation screening in \ac{NSCLC} is essential to determine eligibility for targeted therapies in late stage \ac{NSCLC}~\citep{egfr_nsclc}. The oncogenic status of \ac{EGFR} mutation was determined based on OncoKB annotation~\citep{chakravarty2017oncokb}. \Ac{EGFR} mutations with any oncogenic effect (including predicted/likely oncogenic) were defined as positive label, and \ac{EGFR} mutation with unknown oncogenic status were excluded.

For weakly supervised biomarker prediction, we used Agata~\citep{raciti2022clinical}, as in Sec.~\ref{sec:methods_pancancer}, to transform a set of tiles extracted from \acp{WSI} that belong to to the same case to case-level target labels. 

Virchow is used to generate tile level embeddings on all the evaluated datasets with 224$\times$224 resolution at 20$\times$ magnification. The aggregator networks were trained on the training sets. We selected the best aggregator model based on the \ac{AUROC} score on the validation sets.
Due to the differences between datasets and compared models, we performed a grid search for the initial learning rate of the aggregator training among $1\times10^{-3}$, $1\times10^{-4}$ and $3\times10^{-5}$ and report the best observed test \ac{AUROC} scores in Tab.~\ref{tab:biomarker_results}.

\subsection{Tile-level benchmarking}
\label{sec:methods_tile}
\subsubsection{Linear probing protocol}
For each experiment, we trained a linear tile classifier with a batch size of 4,096 using the \ac{SGD} optimizer with a cosine learning rate schedule, from 0.01 to 0, for 12500 iterations, on top of embeddings generated by a frozen encoder. All embeddings were normalized by z-scoring before classification. Linear probing experiments did not use data augmentation.

\subsubsection{Dataset description}
Dataset details, including training, validation, and testing splits,  are listed in Tab.~\ref{tab:public-tile-datasets}.

\begin{table*}
\centering
\begin{tabular}{llrrccr}
\toprule
Dataset         & Tissue & High-Level Tissue Types & Classes   & Res   & Tile size    & No. of tiles    \\ 
\midrule
PCam            & Lymph node   & 1    & 2         & 10$\times$   & 96$\times$96      & 327,680       \\
WILDS & Lymph node & 1 & 2 & 10$\times$ & 96$\times$96 & 455,954 \\
CRC             & Colon    & 1    & 9         & 20$\times$   & 224$\times$224    & 107,180       \\
CRC (no norm) & Colon    & 1    & 9         & 20$\times$   & 224$\times$224    & 107,180       \\
PanMSK         & PanCancer & 17   & 2         & 20$\times$   & 224$\times$224    & 1,196,171     \\
MHIST           & Colon    & 1    & 2         & 5$\times$   & 224$\times$224    & 3,152         \\
\bottomrule
\end{tabular}
\caption{Summary of the tile-level benchmark datasets used for linear probing.}
\label{tab:public-tile-datasets}
\end{table*}

\textbf{PanMSK}. For a comprehensive \ac{ID} benchmark, 3,999 slides across the 17 tissue types in Fig.~\ref{fig:teaser}d were held-out from the training dataset collected from \ac{MSKCC}. Of these, 1,456 contained cancer that was either partially or exhaustively annotated with segmentation masks by board-certified pathologists. These annotations were used to create a tile-level dataset of cancer vs non-cancer classification which we refer to as \textit{PanMSK}. All images in PanMSK are 224$\times$224 pixel tiles at 0.5 \ac{mpp}. See Sec.~\ref{sec:appendix_panMSK} for further details.

\textbf{\acs{CRC}}. The \ac{CRC} classification public dataset~\citep{kather2019predicting} contains 100,000 images (224$\times$224 pixels) at 20$\times$ magnification sorted into nine morphological classes. We performed linear probing with both the Macenko-stain-normalized (NCT-CRC-HE-100K) and unnormalized (NCT-CRC-HE-100K-NONORM) variants of the dataset. It should be noted that the training set is normalized in both cases and only the testing subset is unnormalized in the latter variant. Thus, the unnormalized variant of \ac{CRC} involves a distribution shift from training to testing.

\textbf{WILDS}
The Camelyon17-WILDS dataset is a public dataset comprising 455,954 images, each with a resolution of 96x96 pixels, taken at 10$\times$ magnification and downsampled from 40$\times$. This dataset is derived from the larger Camelyon17 dataset and focuses on lymph node metastases. Each image in the dataset is annotated with a binary label indicating the presence or absence of a tumor within the central 32x32 pixel region. Uniquely designed to test \ac{OOD} generalization, the training set is composed of data from three different hospitals, while the validation and testing sets each originate from separate hospitals not represented in the training data.

\textbf{MHIST}. The colorectal polyp classification public dataset (MHIST,~\citep{wei2021petri}) contains 3,152 images (224$\times$224 pixels) presenting either hyperplastic polyp or sessile serrated adenoma at 5$\times$ magnification (downsampled from 40$\times$ to increase the field of view). 

\textbf{\acs{PCam}}. The \ac{PCam} public dataset consists of 327,680 images (96$\times$96 pixels) at 10$\times$ magnification, downsampled from 40$\times$ to increase the field of view~\citep{veeling2018rotation,bejnordi2017diagnostic}. Images are labeled as either cancer or benign. We upsampled the images to 224$\times$224 pixels to use with Virchow.

\subsubsection{Qualitative feature analysis}
\label{sec:methods_tile_pca}
We performed an unsupervised feature analysis similar to the procedure in \cite{oquab2023dinov2}, using the \ac{consep} dataset~\citep{graham2019hover} of \ac{HE} stained slides with colorectal adenocarcinoma. \ac{consep} provides nuclear annotations of cells in the following 7 categories: normal epithelial, malignant/dysplastic epithelial, fibroblast, muscle, inflammatory, endothelial, and miscellaneous (including necrotic, mitotic, and cells that couldn't be categorized). Since \ac{consep} images are of size 1,000$\times$1,000 and Virchow takes in images of size 224$\times$224, we resized images to 896$\times$896 and divided them into a 4$\times$4 grid of non-overlapping 224$\times$224 sub-images before extracting tile-level features. For a given image, we used \ac{PCA} on all the tile features from the sub-images, normalized the first and second principal components to values within $[0, 1]$, and thresholded at 0.5. Figure~\ref{fig:linear_probing}d shows some examples of the unsupervised feature separation achieved in this way.

\section*{Acknowledgements}
We gratefully thank Philip Rosenfield from Microsoft and Djamilia Dierov from Paige for their contributions in making this collaboration possible, and Wayne Hendricks from Paige for infrastructure support. 

\FloatBarrier
\bibliography{references}

\onecolumn
\vfill
\pagebreak
\FloatBarrier

\appendix

\renewcommand\thefigure{\thesection\arabic{figure}}
\setcounter{figure}{0}
\renewcommand\thetable{\thesection\arabic{table}}
\setcounter{table}{0}

\section{Appendix}
\subsection{Early foundation models in computational pathology}
\label{app:hist_FM}
Several computational pathology models have been released in the past couple of years. Wang et al.~\cite{wang2022transformer} introduced the first such model leveraging data from \ac{TCGA}~\cite{weinstein2013cancer} and \ac{PAIP}~\cite{kim2021paip} data and a modified MoCoV3~\cite{chen2021mocov3} algorithm to train a 28M parameter SwinTransformer model. Since then, several models using \ac{TCGA} and different model architectures and training procedures have been released: Phikon~\cite{filiot2023scaling} a ViT-B 86M parameter model using iBOT~\cite{zhou2021ibot}, Remedis~\cite{azizi2023robust} a ResNet-152 with 232M parameters and Ciga et al.~\cite{ciga2022self} ResNets with 11-45M parameters using SIMCLR~\cite{chen2020simple}, and Lunit~\cite{kang2023benchmarking} a ViT-S 22M parameter model using DINO~\cite{caron2021emerging}. UNI~\cite{chen_general-purpose_2023} and RudolphV~\cite{dippel2024rudolfv} both leverage proprietary datasets of approximately 100k \acp{WSI} to train a ViT-L 307M parameter model using DINOv2~\cite{oquab2023dinov2}. Campanella et al.~\cite{campanella2023computational} also use a proprietary dataset of 400k \acp{WSI}, although they train a smaller ViT-S with 22M parameters using DINO~\cite{caron2021emerging}.

All of the aforementioned models are summarized in Tab.~\ref{tab:hist_FM}.

\begin{table}[h]
\centering
\begin{tabular}{rcrrcrc}
\toprule
\multirow{2}[0]{*}{Model} & \multirow{2}[0]{*}{Data source} & \multicolumn{2}{c}{Data size}      & \multirow{2}[0]{*}{Model architecture} & \multirow{2}[0]{*}{Model size} & \multirow{2}[0]{*}{Objective function} \\
\cmidrule{3-4}
                       &                              & WSI                       & Tiles &                                     &                             &                                     \\ \toprule
Virchow  (This work)              & MSKCC                        & 1.5M &   2B      & ViT-H                               & 632M                        & DINOv2                              \\ \midrule
UNI~\cite{chen_general-purpose_2023}                    & Mass-100K                    & 100K & 100M    & ViT-L                               & 307M                        & DINOv2                              \\ \midrule
RudolfV~\cite{dippel2024rudolfv} & TCGA + Properitary & 103K & 750M & ViT-L & 307M & DINOv2 \\ \midrule
Campanella et al.~\cite{campanella2023computational}       & Mount Sinai                  & 400K & 3B      & ViT-S                               & 22M                         & DINO                                \\ \midrule
Lunit~\cite{kang2023benchmarking}                  & TCGA + TULIP                 & 37K  & 33M     & ViT-S                               & 22M                         & DINO                                \\ \midrule
Phikon~\cite{filiot2023scaling}                 & TCGA                         & 6K   & 43M     & ViT-B                               & 86M                         & iBOT                                \\ \midrule
Remedis~\cite{azizi2023robust}                & TCGA                         & 29K  & 50M     & ResNet-152                          & 232M                        & SIMCLR                              \\ \midrule
Ciga et al.~\cite{ciga2022self}            & TCGA + CPTAC ++              & 25K  & 4.2M    & ResNet                              & 11-45M                      & SIMCLR                              \\ \midrule
Ctranspath~\cite{wang2022transformer}             & TCGA + PAIP                  & 32K  & 15M     & SwinTransformer                     & 28M                         & MoCoV3                              \\ \midrule
HIPT~\cite{chen2022scaling}                   &        TCGA                      & 11K  & 104M    & ViT-HIPT                            & 10M                         & DINO                                \\ \midrule
LongViT~\cite{wang2023image}              & TCGA                         & 10K  & 1M      & LongNet                             & 22M                         & DINO                                \\ \midrule
PLIP~\cite{huang2023visual}                   & OpenPath                     & NA   & 200K    & ViT-B*                              & 86M                         & CLIP                                \\ \midrule
QUILTNet~\cite{ikezogwo2023quilt}               & Quilt-1M                     & NA   & 1M      & ViT-B*                              & 86M                         & CLIP                                \\ \midrule
CONCH~\cite{lu2023towards}                 & PMC-Path + EDU               & NA   & 1.2M    & ViT-B*                              & 86M                         & CLIP                                \\ \bottomrule
\end{tabular}
\caption{Summary of proposed foundation models in computational pathology highlighting the size of the training data, size of the model architecture, and training objective. The last three entries in the table combine vision and language data and train only using tiles. The model architecture in these cases refers only to the tile embedding as opposed to the entire model size.}
\label{tab:hist_FM}
\end{table}

\FloatBarrier
\subsection{Multi-tissue PanMSK dataset}
\label{sec:appendix_panMSK}

Exhaustive annotations (i.e. a complete segmentation of cancer vs non-cancer regions across the entire \ac{WSI}) were collected for 399 prostate slides, 187 breast slides, 115 bladder slides, 64 breast lymph node slides, and 55 colon slides by a different pathologist for each tissue group. For the other tissue groups (see Fig.~\ref{fig:teaser}d), a pathologist highlighted one or more cancer regions on each slide non-exhaustively. The fully-annotated 64 breast lymph node slides were combined with 48 lymph node slides with highlighted cancer regions, originating from various locations. We sampled non-cancer tiles from slides labeled as benign. With the exception of the endometrial tissue group (for which we selected cancer regions in 11 slides), no tissue group had less than 50 slides partially or thoroughly annotated. We found that when randomly splitting the data into training, validation, and testing subsets, we need at least 30 slides per tissue group to minimize the chance that the training set is not a representative sample of the testing set; therefore, we preferred maximizing slide and patient diversity over maximizing how much of each slide is annotated.

\begin{figure*}[h]
    \centering
    \includegraphics[width=0.95\textwidth]{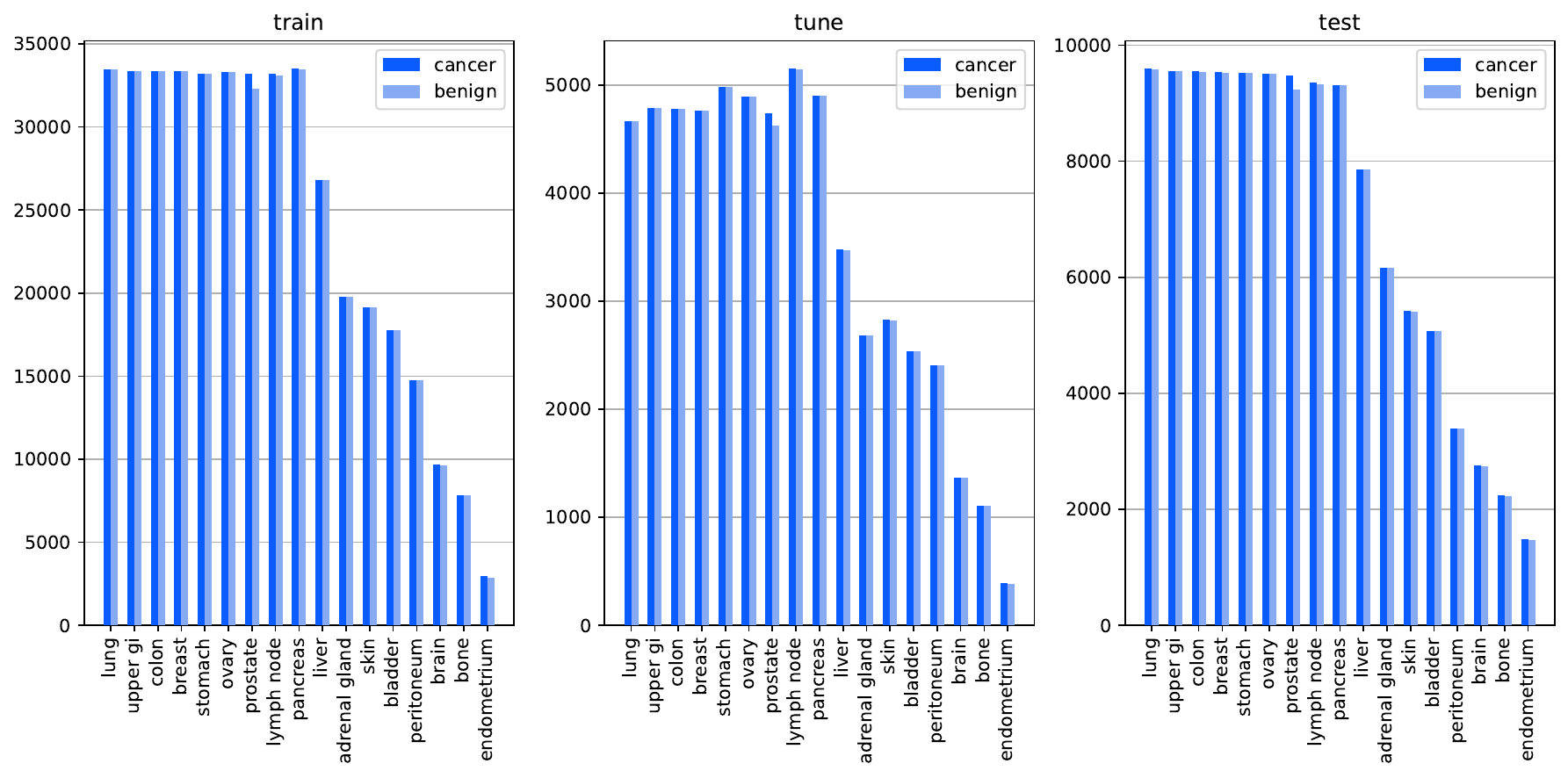}
    \caption{Distributions of cancer and benign tiles in the PanMSK dataset. The splits are balanced such that each tissue group approximately follows the same 7:1:2 (training:validation:testing) ratios in both tiles and slides counts.}
    \label{fig:panmsk_splits}
\end{figure*}

PanMSK was split into training, validation, and testing subsets at the slide level, ensuring that no two subsets share tiles from the same slide. The subsets were balanced to achieve an approximately 7:1:2 ratio of both slides \textit{and} tiles, to equalize the ratio of tile diversity to tile quantity across all splits. The slides were divided into training, validation, and testing subsets by the tissue group and slide-level label (cancer/benign). In order to reduce tissue bias, the number of available cancer tiles for the tissue groups with the most cancer tiles was reduced to the median number of cancer tiles across all tissue groups. The optimal training/validation/testing split was then determined algorithmically by matching the distribution over tissue groups and over labels as closely as possible across all splits. This objective was optimized iteratively. In each iteration, slides were randomly shuffled between the splits and a permutation was picked greedily to maximize the objective. After balancing cancer tiles across the training, validation, and testing subsets and across tissue groups, benign tiles were sampled per tissue group to achieve a 1:1 ratio between cancer and benign tiles. See Fig.~\ref{fig:panmsk_splits} and Tab.~\ref{tab:msk-tile-stats} for more information on PanMSK splits. The exact tile-level data distribution is shown for the training (``train''), validation (``tune''), and testing (``test'') sets of the PanMSK dataset in Fig.~\ref{fig:panmsk_splits}. For each tissue group, sampling of benign and cancerous tiles is balanced. All three splits follow the same data distribution across tissue groups. Virchow performance, stratified by tissue, is shown in Tab.~\ref{tab:per_tissue_panmsk}.

\begin{table}[!htb]
\centering
\begin{tabular}{@{}rrrr@{}}
\toprule
Split        & Slides    & Cancer tiles  & Benign tiles  \\ 
\midrule
Training   & 2,797     & 418,738       & 417,466       \\
Validation     & 402       & 60,462        & 60,296        \\
Testing    & 800       & 119,792       & 119,417       \\ 
\bottomrule
\end{tabular}
\caption{Slide and tile counts in PanMSK. The tiles were split into train, validation, and test subsets with no slide overlap between the subsets. They follow a 7:1:2 split on both the slide- and tile-level.}
\label{tab:msk-tile-stats}
\end{table}

\begin{table}[h]
\centering
\begin{tabular}{rrr} 
 \toprule
 Tissue & Accuracy & F1 Score\\
 \midrule
Adrenal gland & 0.941 & 0.949 \\
Bladder & 0.981 & 0.986 \\
Bone & 0.957 & 0.973 \\
Brain & 0.928 & 0.927 \\
Breast & 0.880 & 0.885 \\
Colon & 0.922 & 0.922 \\
Endometrium & 0.957 & 0.989 \\
Liver & 0.971 & 0.984 \\
Lung & 0.948 & 0.950 \\
Lymph node & 0.928 & 0.935 \\
Ovary & 0.878 & 0.887 \\
Pancreas & 0.970 & 0.973 \\
Peritoneum & 0.935 & 0.943 \\
Prostate & 0.904 & 0.902  \\
Skin & 0.971 & 0.983 \\
Stomach & 0.824 & 0.883 \\
Upper GI & 0.973 & 0.974 \\
\midrule
Overall & 0.975 & 0.927 \\
\bottomrule
\end{tabular}

\caption{Per-tissue tile-level cancer classification performance using Virchow. Overall performance is measured by combining all tiles across all tissues prior to metric computation.}
\label{tab:per_tissue_panmsk}
\end{table}

\FloatBarrier
\subsection{Model training method}
\label{sec:model_arch}
\begin{figure*}[htb!]
    \centering
    \includegraphics[width=0.95\textwidth]{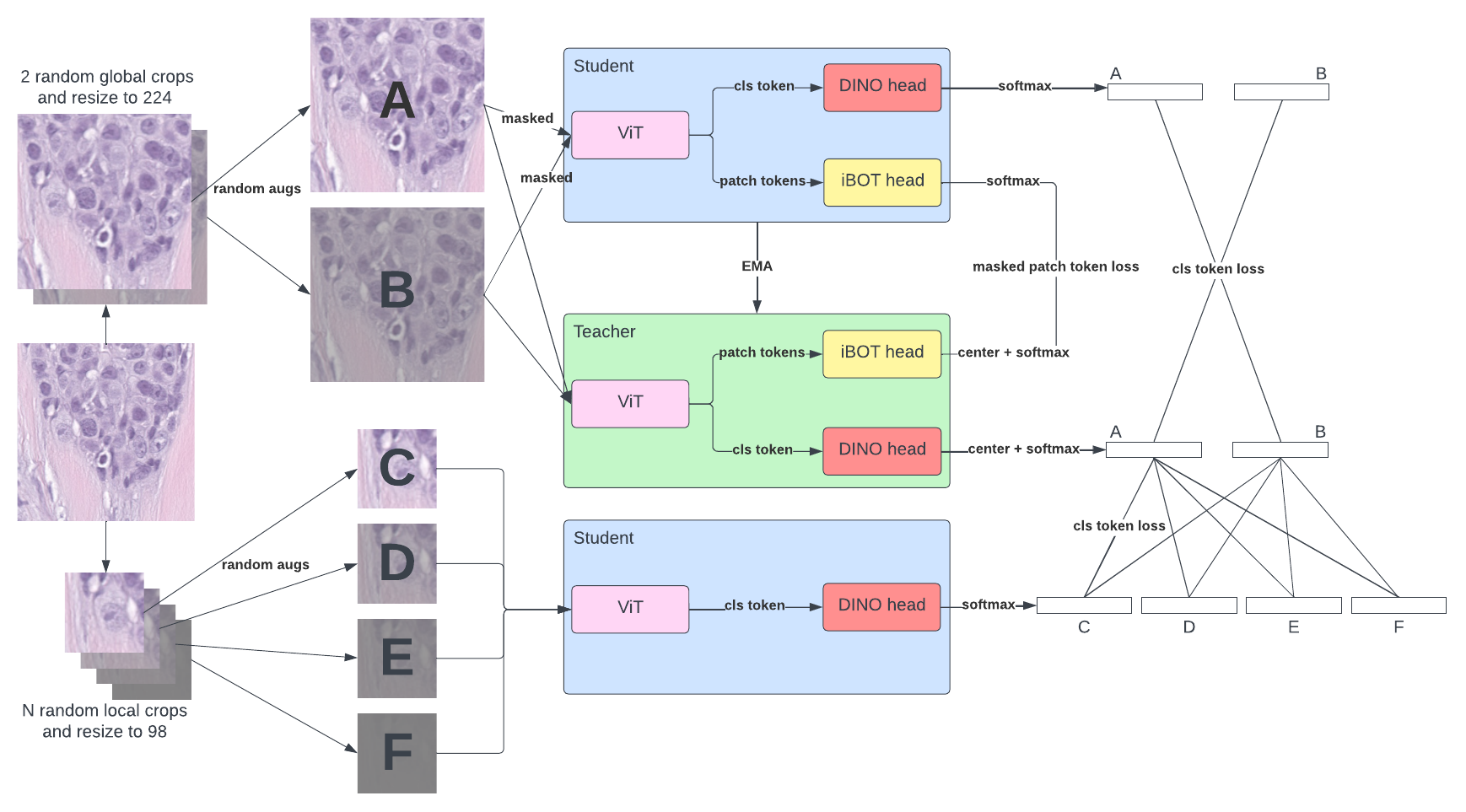}
    \caption{Schematic of DINOv2 training routine. From a single tile, 2 global crops and 8 local crops all with random augmentations are created. The global crops are randomly masked and fed to the student model while the unmasked versions are fed to the teacher model. The student tries to produce a global representation of the views (via the cls token) that matches the teacher's representation of the opposite view. The student also tries to produce representations of the masked image tokens that match the teacher's representations of the same tokens but unmasked. The local crops are only fed to the student which tries to produce a representation that matches the teacher's representations of the global crops. The teacher is an \ac{EMA} copy of the student.}
    \label{fig:dinov2}
\end{figure*}

An overview of the self-supervised Dino v2 training method is shown in Fig.~\ref{fig:dinov2}. Virchow used a ViT-H architecture, trained with Dino v2.

\FloatBarrier
\subsection{Pan-cancer aggregator architecture details}
\label{sec:agata-details}
The Agata aggregator learns to attend to tiles that contribute toward the label decision using cross-attention. The operation is defined using query $Q$, key $K$, and value matrix $V$:
$\text{softmax}\mleft( QK^T / \sqrt{d_k} \mright) V$, where $d_k$ is the output dimension of the key matrix.
In contrast to the typical cross attention mechanism where $Q,K,V$ are projected from the inputs, $Q$ is parameterized directly by the model to reduce GPU memory consumption. $K$ and $V$ are obtained with two consecutive \ac{GELU}~\citep{hendrycks2016gelu} projection layers as: $K = \text{GELU}(W_1^T x + b_1),
V = \text{GELU}(W_2^T K + b_2)$, 

where $x$ is the tile embedding, and $W_n, b_n$ are the weight and bias parameters for the projection layers. In our experiments, $W_1$ produces 256-dimensional keys, $W_2$ produces 512-dimensional values, and we omit scaling by $\sqrt{d_k} = 16$. After the attention step, a sequence of linear layers with non-linear activation (ReLU) are used followed by a final linear layer with softmax activation.

\begin{figure*}
    \centering
    \includegraphics[width=0.9\linewidth]{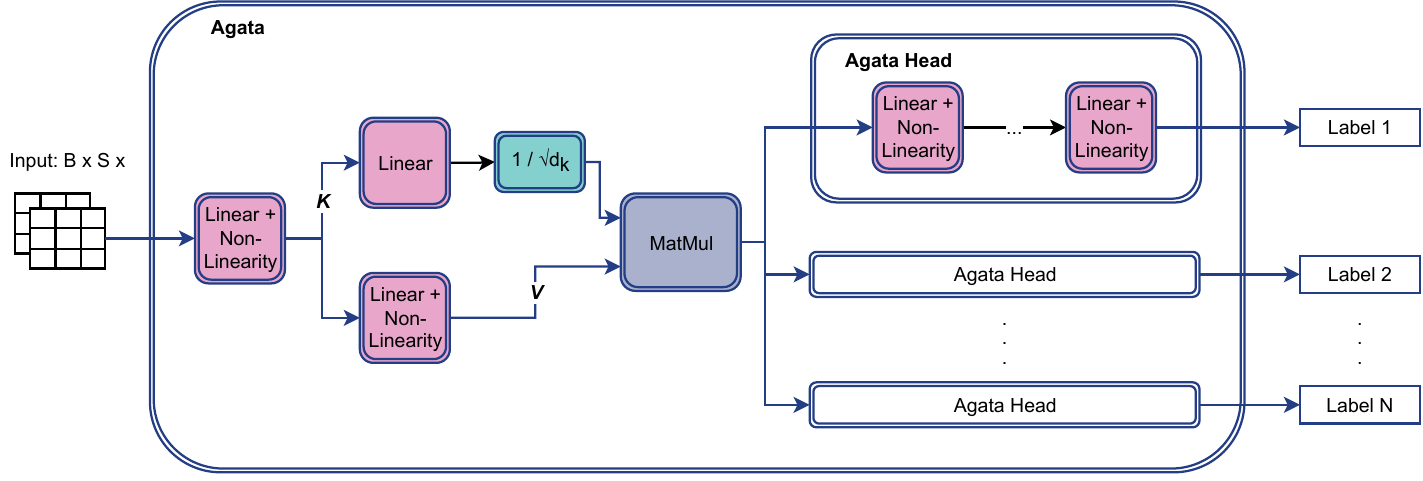}
    \caption{Agata architecture used for specimen-level pan-cancer detection (Sec.~\ref{sec:methods_pancancer}) and biomarker detection tasks (Sec.~\ref{sec:biomarker})}
    \label{fig:agata}
\end{figure*}

\FloatBarrier
\subsection{Tile-level benchmarks}

Additional evaluation metrics for each model on the tile-level benchmarks are detailed in Tab.~\ref{tab:tile_bench}. We report accuracy, balanced accuracy, and weighted F1 score. Balanced accuracy is calculated by averaging \ac{TPR} ($\text{TPR}=\frac{\text{\acs{TP}}}{\text{\acs{TP}}+\text{\acs{FN}}}$) and \ac{TNR} ($\text{\acs{TNR}}=\frac{\text{\acs{TN}}}{\text{\acs{TN}}+\text{\acs{FP}}}$). Weighted F1 score is calculated by first calculating the F1 score (harmonic mean of precision and recall) for each class and then averaging the scores, weighted by the number of positive samples for each class. For balanced accuracy and weighted F1 score calculation, we use the probability $\text{threshold}=0.5$ as the operating point.

\begin{table*}[htbp]
  \centering
    \resizebox{\textwidth}{!}{\begin{tabular}{rrrrrrrrr}
    \toprule
         Dataset & Metric & NatImg & PLIP & CTransPath & $\text{DINO}_{p=8}$ & Phikon & Virchow \\
    \midrule
    \multirow{3}[0]{*}{PanMSK}
          & Accuracy          & 0.883 & 0.862 & 0.897 & 0.903 & 0.924 & 0.950 \\
          & Balanced Accuracy & 0.883 & 0.862 & 0.897 & 0.903 & 0.924 & 0.950 \\
          & Weighted F1       & 0.883 & 0.862 & 0.897 & 0.903 & 0.923 & 0.950 \\
    \midrule
    \multirow{3}[0]{*}{CRC}
          & Accuracy          & 0.952 & 0.946 & 0.962 & 0.959 & 0.958 & 0.973 \\
          & Balanced Accuracy & 0.926 & 0.918 & 0.947 & 0.945 & 0.944 & 0.962 \\
          & Weighted F1       & 0.952 & 0.944 & 0.962 & 0.959 & 0.959 & 0.973 \\
    \midrule
    \multirow{3}[0]{*}{CRC (no norm)}
          & Accuracy          & 0.927 & 0.794 & 0.840 & 0.949 & 0.883 & 0.968 \\
          & Balanced Accuracy & 0.894 & 0.742 & 0.825 & 0.919 & 0.872 & 0.960 \\
          & Weighted F1       & 0.927 & 0.806 & 0.844 & 0.949 & 0.888 & 0.968 \\
    \midrule
    \multirow{3}[0]{*}{WILDS}
          & Accuracy          & 0.934 & 0.869 & 0.947 & 0.957 & 0.971 & 0.970 \\
          & Balanced Accuracy & 0.934 & 0.869 & 0.947 & 0.957 & 0.971 & 0.970 \\
          & Weighted F1       & 0.934 & 0.867 & 0.947 & 0.957 & 0.971 & 0.970 \\
    \midrule
    \multirow{3}[0]{*}{PCam}
          & Accuracy          & 0.886 & 0.874 & 0.872 & 0.918 & 0.906 & 0.933 \\
          & Balanced Accuracy & 0.886 & 0.874 & 0.872 & 0.918 & 0.906 & 0.933 \\
          & Weighted F1       & 0.886 & 0.873 & 0.872 & 0.918 & 0.905 & 0.933 \\
    \midrule
    \multirow{3}[0]{*}{MHIST}
          & Accuracy          & 0.826 & 0.801 & 0.817 & 0.771 & 0.795 & 0.834 \\
          & Balanced Accuracy & 0.821 & 0.786 & 0.801 & 0.746 & 0.782 & 0.830 \\
          & Weighted F1       & 0.827 & 0.801 & 0.816 & 0.769 & 0.796 & 0.835 \\
    \bottomrule
    \end{tabular}}%
    \caption{Downstream task linear probing evaluations. Refer to the text for details on the metrics.}
  \label{tab:tile_bench}%
\end{table*}%

\clearpage
\subsection{Acronyms}
\begin{acronym}[MSK-IMPACT]
    \acro{AI}{artificial intelligence}
    \acro{AUROC}[AUC]{area under (the receiver operating characteristic) curve}
    \acro{CDH1}{Cadherin 1}
    \acro{consep}[CoNSeP]{colorectal nuclear segmentation and phenotypes}
    \acro{CRC}{colorectal cancer}
    \acro{dMMR}{deficient mismatch repair}
    \acro{EGFR}{epidermal growth factor receptor}
    \acro{EMA}{exponential moving average}
    \acro{FGFR}{fibroblast growth factor receptor}
    \acro{FN}{false negative}
    \acro{FPR}{false positive rate}
    \acro{FP}{false positive}
    \acro{GELU}{Gaussian Error Linear Unit}
    \acro{GI}{gastrointestinal}
    \acro{HN}[H\&N]{head and neck}
    \acro{HE}[H\&E]{hematoxylin and eosin}
    \acro{HIPT}{hierarchical image pyramid transformer}
    \acro{iBOT}{image BERT pre-training with online tokenizer}
    \acro{ID}{in-distribution}
    \acro{IHC}{immunohistochemistry}
    \acro{LOH}{loss-of-heterozygosity}
    \acro{MAE}{masked autoencoder}
    \acro{MIL}{multiple instance learning}
    \acro{MMR}{mismatch repair}
    \acro{mpp}{microns-per-pixel}
    \acro{MSI-H}{high-frequency MSI}
    \acro{MSI}{microsatellite instability}
    \acro{MSK-IMPACT}{MSK-Integrated Mutation Profiling of Actionable Targets}
    \acro{MSKCC}{Memorial Sloan Kettering Cancer Center}
    \acro{NSCLC}{non-small cell lung cancer}
    \acro{OOD}{out-of-distribution}
    \acro{PAIP}{Pathology AI Platform}
    \acro{PCam}{PatchCamelyon}
    \acro{PCA}{principal component analysis}
    \acro{PCR}{polymerase chain reaction}
    \acro{ROC}{receiver operating characteristic}
    \acro{SGD}{stochastic gradient descent}
    \acro{TCGA}{The Cancer Genome Atlas}
    \acro{TNR}{true negative rate}
    \acro{TN}{true negative}
    \acro{TPR}{true positive rate}
    \acro{TP}{true positive}
    \acro{ViT}{vision transformer}
    \acro{WSI}{whole slide image}
\end{acronym}
\end{document}